\documentclass[english]{article}
\usepackage[T1]{fontenc}
\usepackage[utf8]{inputenc}
\usepackage{amsmath}
\usepackage{amssymb}
\usepackage{graphicx}
\usepackage{amsfonts,float,commath}
\usepackage{xcolor}

\makeatletter

\usepackage{INTERSPEECH2021}

\makeatother

\usepackage{babel}
\begin{document}
\name{Ritwik Giri $^*$\thanks{$^*$ Equal contribution.}, Shrikant Venkataramani $^*$, Jean-Marc Valin, Umut Isik, Arvindh Krishnaswamy}
\address{Amazon Web Services}
\email{\{ritwikg, shriven, jmvalin, umutisik,  arvindhk\}@amazon.com}

\title{Personalized PercepNet: Real-time, Low-complexity Target Voice Separation and Enhancement}
\maketitle
\begin{abstract}

The presence of multiple talkers in the surrounding environment poses a
difficult challenge for real-time speech communication systems considering
the constraints on network size and complexity.
In this paper, we present Personalized PercepNet, a real-time speech
enhancement model that separates a target speaker from a noisy multi-talker
mixture without compromising on complexity of the recently proposed PercepNet.
To enable speaker-dependent speech enhancement, we first show how we can train
a perceptually motivated speaker embedder network to produce a representative
embedding vector for the given speaker.
Personalized PercepNet uses the target speaker embedding as additional
information to pick out and enhance only the target speaker while suppressing
all other competing sounds. Our experiments show that the proposed model
significantly outperforms PercepNet and other baselines, both in terms of
objective speech enhancement metrics and human opinion scores.
\end{abstract}

\section{Introduction}
\label{sec:intro}

With the ubiquitous presence of real-time audio communication systems,
there has been a significant interest in speech enhancement
algorithms that operate in real-time with low complexity.
In the real world, a user (or target speaker) of these communication systems
often finds themselves in the presence of competing background sounds. The goal
of speech enhancement is to extract a high-quality version of a target
speaker's utterance from the mixture that contains the target speaker in
addition to multiple competing ambient sounds. Considering the complexity of
enhancing fullband ($48$ kHz) speech mixtures, a perceptually motivated,
low-complexity model called ``PercepNet'' has been recently shown to deliver
high-quality speech enhancement in real-time even while operating on less than
$5$\% of a CPU core~\cite{valin2020perceptually}.

In more challenging situations, the interference can also include other
speech sources, such as a television, children playing
and other people conversing in the background. Since interfering speech sources
are spectrally similar to the target talker, they are usually not suppressed
by speech enhancement algorithms.
This shortcoming can be addressed through
single-channel multi-talker source separation algorithms that extract and
separate all the speech-like sounds from the given mixture~\cite{luo2018tasnet,
hershey2016deepclustering, kolbaek2017multitalker, luo2019convtasnet,
tzinis2019unsupervised}.
% JMV: I think we need to better emphasize our contribution compared to SOTA
Yet, these source separation approaches do not focus
specifically on extracting the target speaker alone and might make the target speaker's
 signal available on any of the output
channels, bringing the need for additional speaker
tracking~\cite{neumann2019allneural, raj2020integration} which comes with the
cost of increasing network complexity.

Resolving this issue involves a joint approach where the speech enhancement
algorithm is capable of two tasks: (i) identifying the target speaker amidst
all the interfering sounds in the given mixture, and (ii) isolating and enhancing
only the target speaker. To this end, Wang \textit{et al.} proposed
Voicefilter, which performs targeted voice
separation~\cite{muckenhirm2019voicefilter}. To identify the target speaker,
Voicefilter uses a pre-trained speaker embedding network that learns a
discriminative speaker representation from the Mel spectrogram of an audio
signal~\cite{wan2018generalized}. These embeddings are then used to condition
the separation network and isolate only the target speaker. Several other
approaches have also been proposed in recent times~\cite{zmolikova2017speaker, mun2020sound,
gu2019neural, li2020atss, wang2019speakerinventory}. Despite the availability of
several such algorithms, these approaches have primarily focused on target source
separation for non-real-time applications. The use of bidirectional recurrent
layers and large convolutional layers increases the complexity of these models.
Morover, the non-causal nature of the convolutions and the bidirectional recurrent
units makes these aforementioned approaches unsuitable for real-time, low-complexity 
applications. Recently, Voicefilter-lite, a real-time alternative to
the Voicefilter has been proposed~\cite{Wang2020VoiceFilterLiteST} to improve
the performance of speech recognition systems in multi-talker situations. Although
Voicefilter-lite showed impressive performance for overlapped speech recognition,
it was not designed to improve human perception or intelligibility under such
conditions, which is the need of the hour for real-time audio communication systems.

In this paper, we introduce Personalized PercepNet~(Section~\ref{sec:percepnet}),
a perceptually motivated
approach to real-time, low-complexity target speaker enhancement. We
improve upon the speech enhancement capabilities of PercepNet by conditioning
on the target speaker's voice~(Section~\ref{sec:personalized_percepnet}).
This enables PercepNet to distinctly identify
and enhance the target speaker's utterance while suppressing all the other
interferences, even in the presence of multiple talkers or other speech-like
sounds.
Given an audio example of the target speaker's voice, we first compute (offline) a
discriminative embedding representation that captures the identity of the
speaker and distinguishes the target speaker from other speakers.
We then use the computed embedding as additional information to
the separation neural network and extract only the target speaker's voice from
any given mixture. Like in PercepNet, our neural networks operate
on a perceptually motivated feature representation. The features include
perceptually relevant parameters like the spectral envelope and the signal
periodicity, and allow us to operate on a compact $68$-dimensional feature space.
We demonstrate through
our experiments~(Section~\ref{sec:experimental_setup}) that our approach leads
to superior speech enhancement in noisy multi-talker situations both in terms of subjective
listening tests and in terms of objective evaluation metrics~(Section~\ref{sec:evaluation_results}).

\section{PercepNet: An Overview}
\label{sec:percepnet}

The PercepNet algorithm operates on $10$\nobreakdash-ms frames with $30$ ms of look-ahead and 
enhances $48$ kHz speech in real-time. Despite its complexity being much lower
than the maximum allowed by the recently concluded first DNS 
challenge~\cite{reddy2020interspeech}, PercepNet ranked second in the real-time 
track. 

The key elements of the algorithm are (i) a perceptual band 
representation as the feature space, (ii) a perceptually motivated pitch-filter 
and (iii) an RNN model to estimate band ratio masks. 

\noindent\textbf{Feature Space:} Instead of operating on Fourier transform bins 
(like many other speech enhancement methods), PercepNet operates on only 
$32$~triangular spectral bands, spaced according to the equivalent rectangular 
bandwidth (ERB) scale.  The input features used by PercepNet are tied to these 
$32$~ERB bands. For each band, we use two features: the magnitude of the band 
and the pitch coherence (frequency-dependent voicing). We also include
$4$~general features (including the pitch period), resulting in a 
$68$~dimensional feature space.

\noindent \textbf{Pitch Filter:} To reconstruct the harmonic properties of the 
clean speech from the spectral envelopes, PercepNet also employs a comb filter 
controlled by the pitch frequency. Such a time-domain comb filter allows
a much finer frequency resolution than would otherwise be possible with 
the STFT ($50$~Hz using $20$\nobreakdash-ms windows). The comb filter's effect 
is independently controlled in each band using \textit{pitch-filter strength} 
parameters~\cite{valin2020perceptually}.

\noindent\textbf{Model:} PercepNet uses a recurrent neural network (RNN) to 
estimate a ratio mask in each band. This ratio mask can also be 
interpreted as the corresponding gain that needs to be applied to the noisy 
signal to match the clean target's spectral envelope. Along with gains, our 
model also outputs the estimated pitch-filter strength for each band and a 
frame-level Voice Activity Detector (VAD) output.

\section{Personalized PercepNet}
\label{sec:personalized_percepnet}
\begin{figure}[ht!]
	
\centering{\includegraphics[width=1\columnwidth]{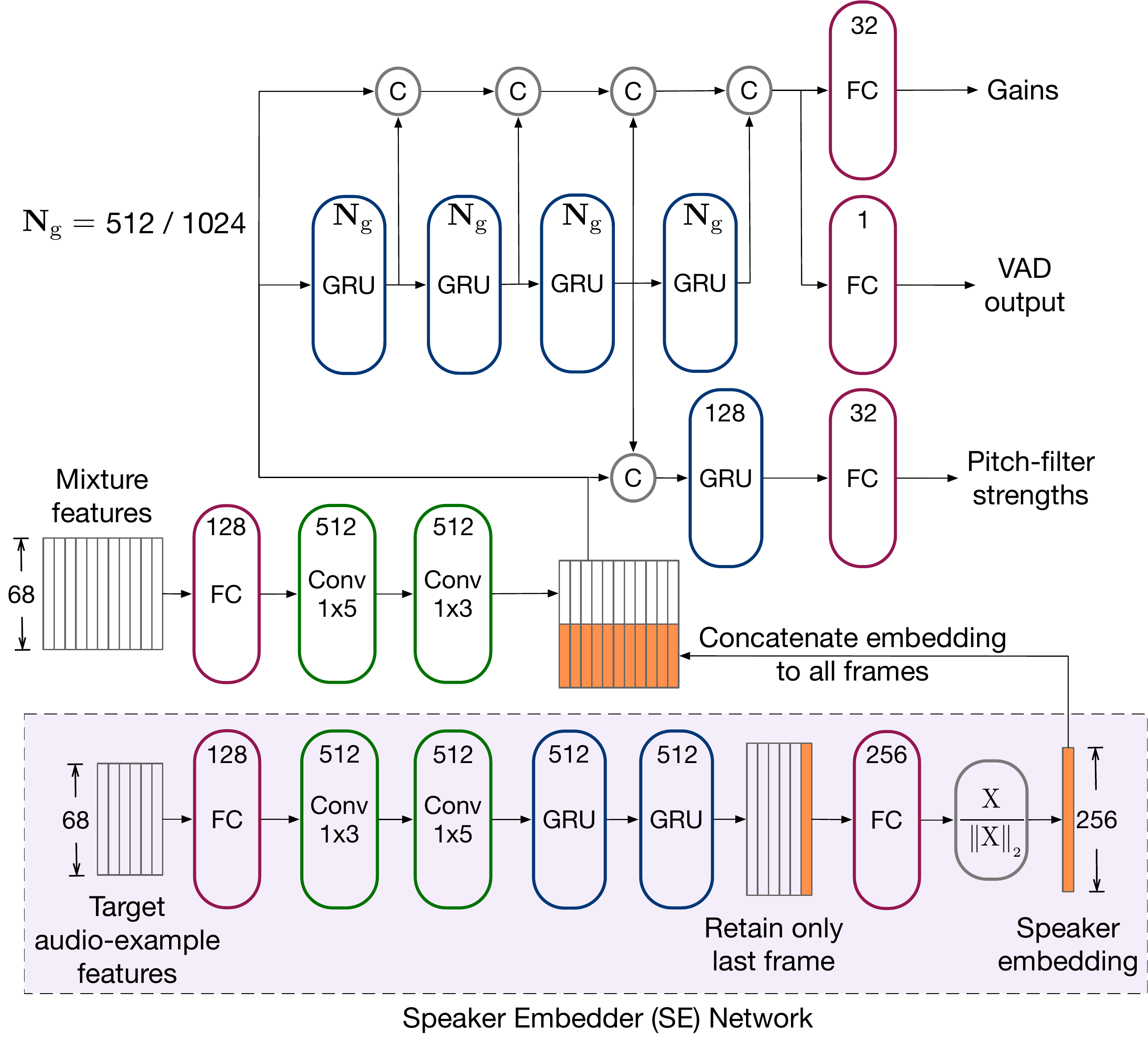}}
	\caption{Block diagram of the speaker-conditioned Deep Neural Network 
(DNN) model used for Personalized PercepNet. The target audio example is used 
to compute the normalized speaker embedding which is appended as additional 
information to the DNN model. The DNN model processes the feature 
representation of the mixture signal and estimates the target speaker filter 
gains and pitch filter strengths at its output. We also train the DNN model to 
estimate the target voice activity (VAD output). The gains and the pitch-filter 
strengths are $32$~dimensional, and the VAD output is a $1$~dimensional output 
signal. The number of units is given above each layer.
All the convolutional layers perform $1$~dimensional convolutions 
along the time dimension. The circles with ``C'' inside denote the 
concatenation of the two inputs arriving at that point. We experiment with
GRUs with either $N_g=512$ or $1024$ units. We refer to these 
models as PPN-$512$ and PPN-$1024$, respectively.}
	\label{fig:speaker_conditioned_dnn}
	\vspace{-1.5em}
\end{figure}

Fig.~\ref{fig:speaker_conditioned_dnn} gives the block diagram of the neural 
network model used for Personalized PercepNet. To identify the target speaker 
in a given mixture, we assume that we have access to an audio example of the 
target speaker's voice during inference. We pre-train a speaker 
verification network that can capture a speaker's identity from a given 
utterance in the form of a representative speaker embedding. That network is
trained once, and then used for any utterance from any target speaker.
The target speaker's embedding is then used by Personalized PercepNet to
distinguish the target speaker from other talkers.

\subsection{Learning Speaker Embeddings}
\label{subsec:speaker_embeddings}
To learn the embedding representation for a target speaker, we extend the work 
in~\cite{wan2018generalized} to train a speaker verification network. The 
underlying goal of speaker verification is to identify whether a given speech 
example belongs to a particular speaker. In doing so, speaker verification 
networks have been shown to learn suitable speaker-discriminative embedding 
representations that have been used for several tasks like target speaker 
diarization~\cite{zhang2019fully}, text-to-speech systems that 
generate outputs in different target voices~\cite{jia2018advances}, voice style 
transfer~\cite{qian2019autovc} and targeted source 
separation~\cite{muckenhirm2019voicefilter}.
We train our \textit{Speaker Embedder} (SE) network to operate on the 
same set of features as PercepNet. The audio example
is converted into a feature representation and sent to the SE 
network. As shown in Fig.~\ref{fig:speaker_conditioned_dnn}, the last frame 
of the final GRU's output is normalized and chosen to be the corresponding 
speaker embedding. To train our SE network, we use the generalized end-to-end 
loss-function described in~\cite{wan2018generalized}.

SE networks are generally trained on full
spectrograms or high-resolution Mel spectrograms to learn discriminative speaker 
embeddings. Instead, we choose to learn speaker 
embeddings from the more compact $68$-dimensional feature 
representation described earlier. One reason why a high resolution
representation is unnecessary is the fact that the pitch period for each frame
is explicitly included as a feature, rather than having to be implicitly 
extracted from the spectrum by the embedding network. In addition, the
LPCNet~\cite{valin2019lpcnet} vocoder has previously demonstrated that
clean speech can be reconstructed with sufficiently high quality based
only on an $18$-band wideband representation, plus pitch and voicing information.

\subsection{Speaker-conditioned DNN}
\label{subsec:speaker_conditioned_dnn}
As seen from Fig.~\ref{fig:speaker_conditioned_dnn}, the input to the DNN 
model is the featurized representation of a speech mixture that contains the 
target speaker in the presence of concurrent interfering talkers and ambient 
noise. We use the SE network and the given clean audio example to obtain an embedding 
representation for the target speaker we wish to isolate. The speaker embedding 
is then appended to every frame before the GRU layers.

\subsection{Loss function}
\label{subsec:loss_function}
To train the DNN model, we reuse the gain and pitch strength loss functions from 
the original PercepNet~\cite{valin2020perceptually}.
We also provide additional supervision in terms of the voice activity of the 
target speaker, as shown in Fig~\ref{fig:speaker_conditioned_dnn}. The
VAD output is expected to produce a value of $1$ for frames 
where the target speaker is active and produce a value of $0$ otherwise. We 
treat the VAD as a binary classification problem and minimize the binary 
cross-entropy between the VAD output and the target VAD label. In 
Fig.~\ref{fig:vad_out}, we demonstrate that the VAD also operates in a 
personalized manner and can identify frames where the target speaker 
is active. 

\section{Experimental Setup }
\label{sec:experimental_setup}
To evaluate the performance of the proposed approach, we compare the 
performance of Personalized PercepNet to that of a PercepNet baseline 
model~\cite{valin2020perceptually} and to the NSNET-$2$ 
model~\cite{braun2020data}. The NSNET-$2$ model has been used as the baseline 
for the Microsoft-organized ICASSP $2021$ Personalized deep noise suppression 
challenge (track 2)~\cite{reddy2020icassp}. We perform this comparison on a 
speech enhancement task where the goal is to extract the target speaker from a 
mixture that contains background noise and an interfering talker. We evaluate 
and compare the speech quality and performance of these models using 
mean-opinion-score (MOS)~\cite{P.800} numbers obtained from subjective listening tests and 
objective evaluation metrics.

\subsection{Training data}
\textbf{Speaker Embedder:} We train the SE network using the original 
VoxCeleb$2$~\cite{chung2018voxceleb2} training set and validate the trained model 
using  VoxCeleb$1$~\cite{nagrani2017voxceleb} test set. 
We use $6$\nobreakdash-sec long inputs that are cropped 
from concatenated utterances for each speaker. We do not use any other 
augmentation to train the SE.

To evaluate the 
effectiveness of the speaker embeddings, we compute the Equal Error Rate (EER) 
on a text-independent speaker verification task as described 
in~\cite{wan2018generalized}. Our SE model -- trained on the same
$68$-dimensional band feature space as the enhancer -- achieves $4.8$\% EER on the 
VoxCeleb$1$ test set~\cite{nagrani2017voxceleb}.

\vspace{0.5em}

\noindent\textbf{DNN model:} We use LibriSpeech~\cite{panayotov2015librispeech}, VoxCeleb$1$~\cite{nagrani2017voxceleb} and VoxCeleb$2$~\cite{chung2018voxceleb2} to 
train the DNN model. For
LibriSpeech, we use the training and development sets as defined in the dataset 
protocol: the training set contains $2338$~speakers, and the development set 
includes $73$~speakers.  A portion of LibriSpeech, specifically 
``train-other-$500$'', has some stationary background noise in it, which makes it 
unusable to train our enhancer as it is. We use a VAD and lightweight denoiser  
(SpeexDSP\footnote{https://gitlab.xiph.org/xiph/speexdsp/})  to eliminate the 
stationary noise before using this data for training. Likewise, the two VoxCeleb 
datasets are collected from television broadcasts and contain 
background
music and other effects. Some of the collected data is also highly reverberant. Following the 
data filtering technique described in~\cite{isik2020poconet}, we isolate the 
clean speech in VoxCeleb$2$ and VoxCeleb$1$ and eliminate reverberant
clips. Thereafter, we include the processed clean-speech clips to the DNN training data 
only if the corresponding speaker has more than $100$ utterances. With these 
steps, we end up with $4500$ distinct speakers.  We use the same noise data 
used in~\cite{valin2020perceptually} that includes $80$ hours of various noise 
types, sampled at $48$ kHz.

We train the DNN model on synthetic mixtures containing the target talker (signal), 
an interfering talker (interference) and noise. These mixtures are generated with signal to noise ratios (SNR) 
ranging from $-5$~dB to $35$~dB and signal to 
interference ratios (SIR) 
ranging from  $-5$~dB to $10$~dB.  To ensure robustness in reverberated conditions, the noisy signal is convolved with 
simulated and measured room impulse responses. To improve the quality of the 
perceived speech the target is set to include the early reflections and only 
attenuate the late reverberations~\cite{valin2020perceptually, 
isik2020poconet}. We improve the generalization of the model by using an 
extensive augmentation stack that includes a low pass filter with a random cut 
off frequency between $3$ kHz and $20$ kHz, and a spectral tilt to simulate 
different microphone frequency responses. 

\subsection{Evaluation data}
For our experiments, we construct a synthetic evaluation set using LibriSpeech 
dev set following~\cite{muckenhirm2019voicefilter}. The only difference 
from~\cite{muckenhirm2019voicefilter} is that we also add background noise to 
the mixture of two speakers (primary and secondary). We use noise clips from the 
DEMAND database~\cite{joachim2013demand}. This ensures that the speech and noise 
examples used for evaluation are completely separate from the training data. 
The interfering talker and noise are set to have SIR and SNR values uniformly 
distributed in the range $15$~dB to $3$~dB. This is done because in 
our applications of interest, the target speaker typically close to the microphone 
and is the loudest component of the mixture. We use $20$\nobreakdash-sec long utterances 
for the mixtures and the target example files and generate $500$~noisy recordings 
for the evaluation set.

\subsection{Performance Metrics}
For subjective testing, we use the ITU-T P.$808$ crowdsourcing approach~\cite{P.808}.
The models' output is rated by
$10$~listeners for each of the $500$~noisy recordings and averaged to produce the MOS scores.
For objective evaluation, we use
wideband PESQ~\cite{P.862} and the 
composite CSIG, CBAK, and COVL scores proposed in~\cite{loizou2007speech}.

\section{Evaluation Results}
\label{sec:evaluation_results}
We consider two versions of the
Personalized PercepNet network shown in Fig.~\ref{fig:speaker_conditioned_dnn}: 
$\mathbf{N}_{\text{g}} = 512$ and $\mathbf{N}_{\text{g}} = 1024$. We refer to these models as 
PPN-$512$ and PPN-$1024$ respectively. Table~\ref{tableobjectivemetrics} and 
Table~\ref{tablesubjectivemetrics} show the objective metrics and the 
subjective listening test scores on the test set for all four 
models. Our results indicate that the Personalized PercepNet models 
significantly outperform the baseline PercepNet and NSNET-$2$ models in objective 
and subjective metrics. The PPN-$1024$ model further improves upon the 
performance of the PPN-$512$ model. These improvements are consistently 
observed in both the mean opinion scores obtained in the listening tests and 
the objective evaluation metrics. 

The  complexity of Personalized PercepNet is mostly dictated by the the number of parameters in the DNN model. The PPN-$512$ model has 8.5M parameters, whereas PPN-$1024$ has $26.5$M parameters. With a $10$\nobreakdash-ms frame size, PPN-$512$ requires $4.7$\% and PPN-$1024$ requires $17.2$\% of one mobile x$86$ core
($1.8$~GHz Intel i$7$\nobreakdash-$8565$U CPU) for real-time operation.

\begin{figure}[h]
	\begin{center}
	\begin{minipage}[b]{\linewidth}
		\centering
		
\centerline{\includegraphics[width=\textwidth]{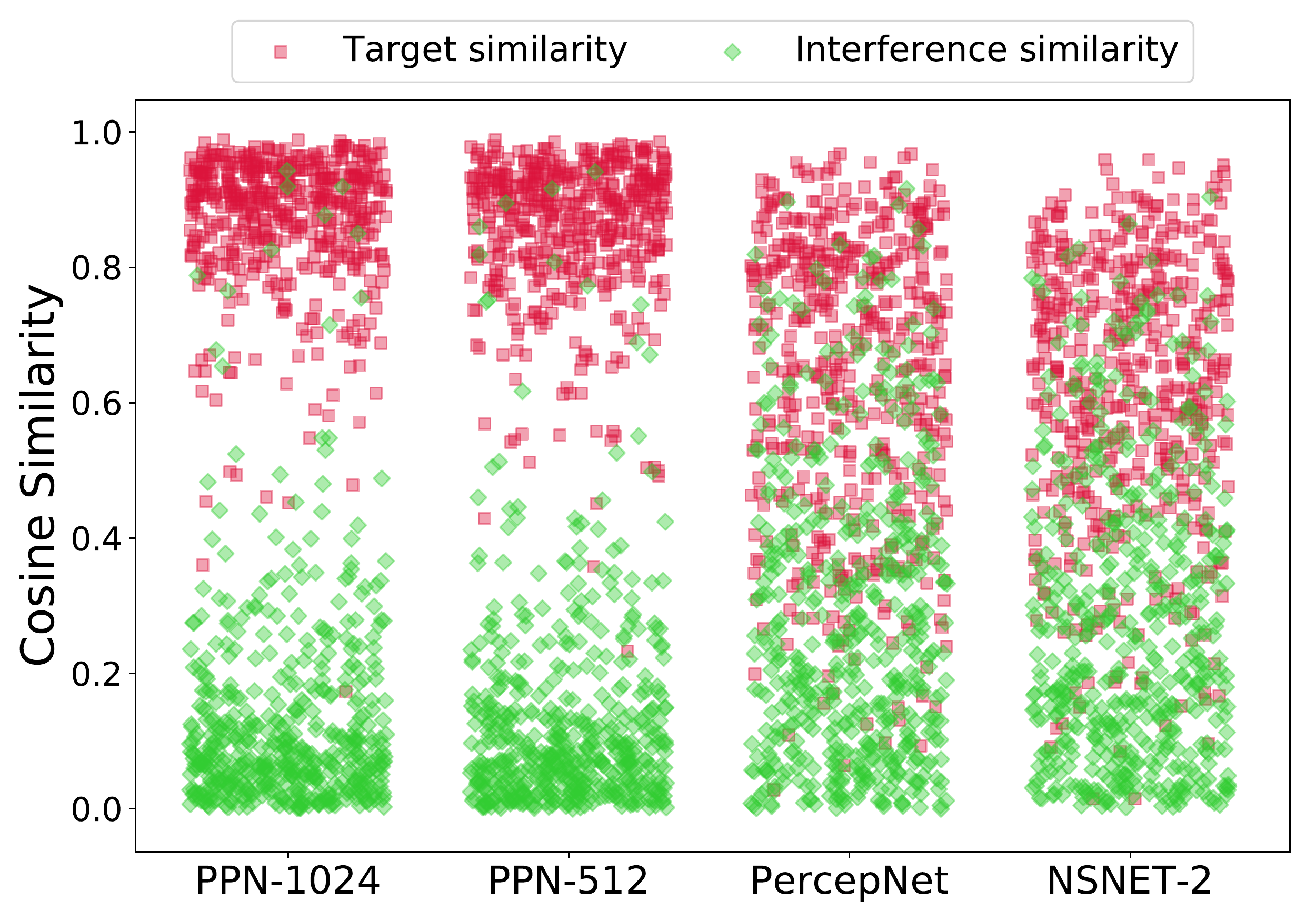}}
	\end{minipage}
	
	\caption{Cosine similarity between model output embedding and 
target/interference embedding. We add random jitter along the horizontal 
axis to the scatter points to improve the readability of the scatter plot. We 
see that the personalized models PPN-$1024$ and PPN-$512$ generally produce 
cleaner outputs containing only the target-speaker's voice.}
	\label{fig:cos_sim}
	\end{center}
\end{figure}

\begin{figure}[h]
	\begin{center}
	\begin{minipage}[b]{\linewidth}
		\centering
		
\centerline{\includegraphics[width=\textwidth]{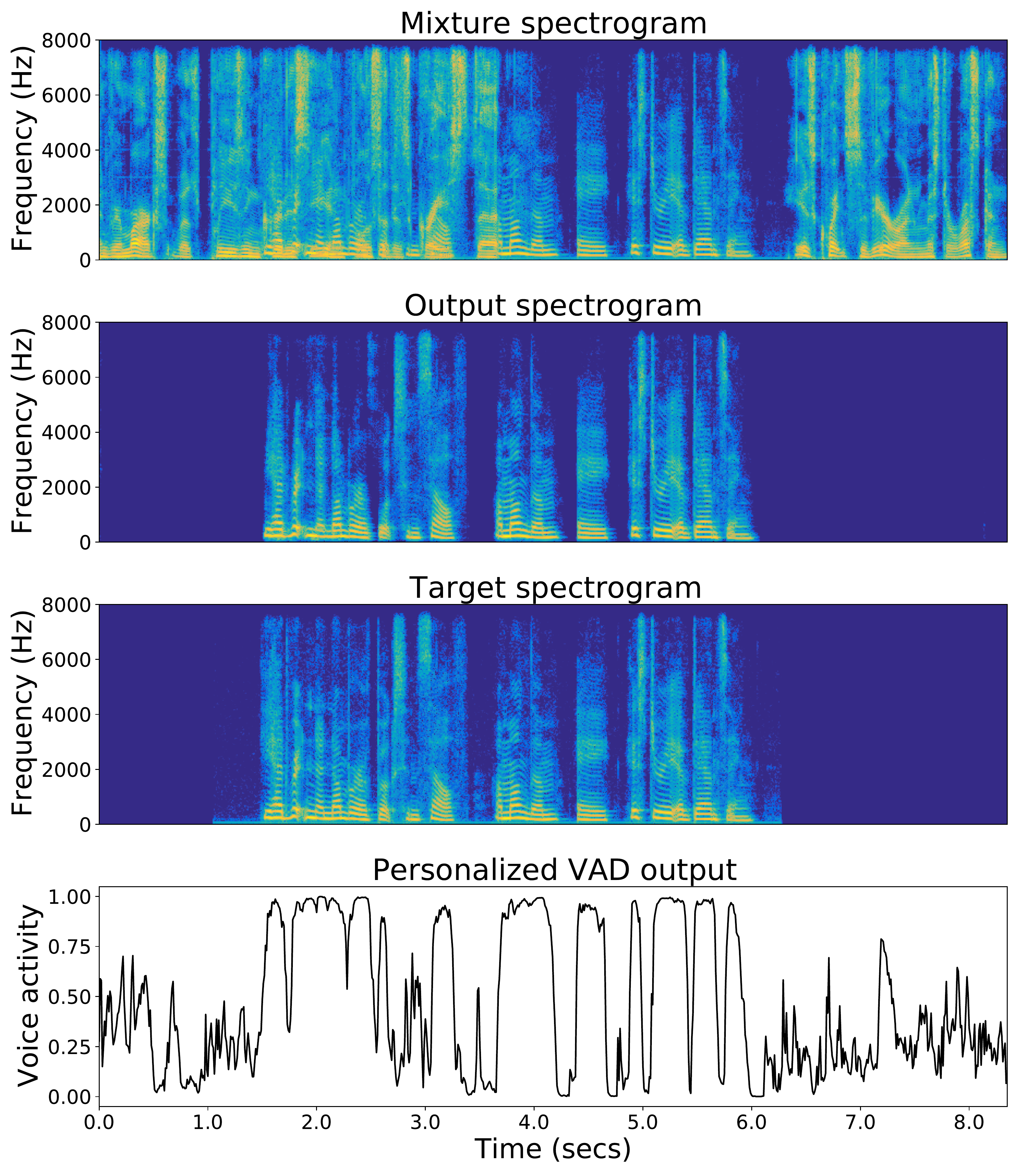}}
	\end{minipage}
	
	\caption{Personalized VAD Output. Although we operate on full-band mixtures and produce full-band outputs, we show spectrograms 
downsampled to $16$ kHz for readability. The VAD output 
indicates speaker activity on a frame by frame basis where there is a frame 
transition every $10$ ms. We transform the frame indices into their equivalent 
time values.}
	\label{fig:vad_out}
	\end{center}
	\vspace{-2em}
\end{figure}

\begin{table}[hbt!]
	\vspace{-0.1in}
	\footnotesize
	\centering
	\caption{Objective evaluation of different algorithms over the 
synthetic test set created from LibriSpeech 
	}
	\begin{tabular}{l|cccc} 
		\hline \\[-2ex]

		Methods  &{PESQ} &{CBAK} &{COVL} &{CSIG}  \\
		\hline
		
		Noisy 
		& 1.455 & 1.877 & 1.886 & 2.482  \\
		NSNET-2   \cite{braun2020data}    &
		1.629 & 2.143 & 1.981 & 2.476  \\
		PercepNet \cite{valin2020perceptually} & 1.748 & 1.989 & 2.052 
& 2.491  \\
		\bf{PPN-512} & 2.357 & 2.491 & 2.871 & 3.462  \\
		\bf{PPN-1024} & \bf{2.412} & \bf{2.528} & \bf{2.920} & 
\bf{3.501} \\
		\hline 
	\end{tabular}
	\vspace{0.05in}
	
	\label{tableobjectivemetrics}

\end{table}

\begin{table}[hbt!]
	
	\footnotesize
	\centering
	\caption{Subjective evaluation (MOS) of different Algorithms over the 
synthetic test set created from LibriSpeech 
	}
	\begin{tabular}{l|c} 
		\hline \\[-2ex]
		
		Methods  &{MOS}  \\
		\hline
		
		Noisy
		&  2.384 \\ 
		NSNET-2   \cite{braun2020data}    & 2.541 \\ 
		PercepNet \cite{valin2020perceptually} & 2.624 \\ 
		\bf{PPN-512} & 3.128  \\ 
		\bf{PPN-1024} & \bf{3.208} \\ 
		\hline 
	\end{tabular}
	\vspace{0.05in}
	
	\label{tablesubjectivemetrics}
	
\end{table}

With Personalized PercepNet, we expect that the model output now only contains 
the enhanced target speaker's utterance contained in the original mixture 
signal. To check if this is indeed the case and the output does not enhance the 
wrong speaker or contain a combination of both speakers, we 
probe our Personalized PercepNet models further. We use the pre-trained SE 
network to compute the speaker embedding of the model output. We compare this computed output embedding to the target speaker's embedding and the 
interfering speaker's embedding in terms of cosine similarity values. For 
the target and interference embeddings, we use the ground-truth target and 
ground-truth interference utterances used to generate the noisy mixture itself. 
Fig.~\ref{fig:cos_sim} demonstrates how well our model has learned to address 
this challenge. We plot the results as a scatter plot of cosine similarity values for all the four models 
(PPN-$1024$, PPN-$512$, PercepNet and NSNET-$2$) over all the $500$ mixtures. 
It is evident from Fig.~\ref{fig:cos_sim} that our model has learned to 
extract only the target speaker's voice. This is seen by the fact that the target 
similarity values are clustered closer to $1$ and the interference similarity values are clustered closer to $0$ for both PPN-$1024$ and PPN-$512$ 
models. The target cluster (red circles) is also well-separated from the 
interference cluster (green triangles) for these models. On the other hand, the 
baseline models have significant overlaps between these clusters.

Finally, Fig.~\ref{fig:vad_out} shows how the VAD output head (from 
Fig.~\ref{fig:speaker_conditioned_dnn}) has learned the target speaker 
activity on a toy mixture constructed from the LibriSpeech dev-set. The use of speaker 
conditioning has enabled the model to learn only the target speaker's voice 
activity while ignoring the interfering voice activity. Hence our proposed 
model can also be used as a personal VAD~\cite{ding2019personal}. Further 
experiments on the quality of the VAD output are beyond the scope of this 
work. 

Casual listening confirms that Personalized PercepNet is able to better isolate
the target speaker's voice than the baseline PercepNet. In doing so, one
artifact we sometimes notice is a form of pitch modulation, especially when
the target and interfering talkers overlap. We believe this is due to pitch
estimation errors in the overlap case. While the artifact is usually not
annoying, we believe a better pitch estimator would help further improve
quality.

\section{Conclusion}
In this paper, we present Personalized PercepNet, a real-time speech 
enhancement algorithm that enhances the target speaker from a mixture that 
contains ambient noise and other interfering talkers. The neural network of the 
proposed model consists of two components: the speaker embedder network and the 
DNN model. To train the speaker embedder network, we rely on a relatively 
compact set of $68$ perceptually motivated features like spectral envelopes and 
speech periodicity and learn discriminative speaker embeddings. Using the 
embeddings for the target speaker, the DNN model then operates on the feature 
representation of the mixture and extracts the target speaker. Our experiments 
confirm that the proposed approach improves upon the baseline PercepNet model 
significantly without compromises in real-time operation or memory constraints.

%\newpage
\bibliographystyle{IEEEbib.bst}
\bibliography{refs.bib}

\end{document}